# Eye-gaze Estimation with HEOG and Neck EMG using Deep Neural Networks

Zhen Fu[1], Bo Wang[1], Fei Chen[2], Xihong Wu[1,3], Jing Chen[1,3]

[1]*Department of Machine Intelligence, Speech and Hearing Research Center, and Key Laboratory of Machine Perception (Ministry of Education), Peking University,* Beijing, China
[2]*Department of Electrical and Electronic Engineering, Southern University of Science and Technology,* Shenzhen, China
[3]*Peng Cheng Laboratory,* Shenzhen, China
{fuzhen364, wangbo1351, xhwu, janechenjing}@pku.edu.cn, fchen@sustech.edu.cn

*Abstract*—Hearing-impaired listeners usually have troubles attending target talker in multi-talker scenes, even with hearing aids (HAs). The problem can be solved with eye-gaze steering HAs, which requires listeners eye-gazing on the target. In a situation where head rotates, eye-gaze is subject to both behaviors of saccade and head rotation. However, existing methods of eye-gaze estimation did not work reliably, since the listener's strategy of eye-gaze varies and measurements of the two behaviors were not properly combined. Besides, existing methods were based on hand-craft features, which could overlook some important information. In this paper, a head-fixed and a head-free experiments were conducted. We used horizontal electrooculography (HEOG) and neck electromyography (NEMG), which separately measured saccade and head rotation to commonly estimate eye-gaze. Besides traditional classifier and hand-craft features, deep neural networks (DNN) were introduced to automatically extract features from intact waveforms. Evaluation results showed that when the input was HEOG with inertial measurement unit, the best performance of our proposed DNN classifiers achieved 93.3%; and when HEOG was with NEMG together, the accuracy reached 72.6%, higher than that with HEOG (about 71.0%) or NEMG (about 35.7%) alone. These results indicated the feasibility to estimate eye-gaze with HEOG and NEMG.

I. INTRODUCTION

Hearing-impaired (HI) listeners usually have troubles to attend to the intended talker in a complex auditory scene with multiple simultaneous talkers and background noises [1]. Advanced hearing aids (HAs) could suppress background noises to a certain degree, by utilizing technologies such as noise reduction algorithms and directional microphones (i.e., beamforming) [2]. However, noise reduction algorithms usually need to assume the acoustic distribution of the target and background sounds and this could be problematic in multi-talker scenarios where listeners could switch the attentional target. The beamforming technology is usually with forward-pointing directional microphones or microphone array, which could enhance signals in front of the listener. Nonetheless, the benefit is limited since listeners do not usually orient their heads toward the target sound [3], [4]. In general, HAs usually fail to amplify the target speech stream without amplifying other streams in multi-talker scenarios, because they are unaware of listener's attentional target.

In face-to-face conversational scenes, auditory and visual cues are both available. It is suggested that the visual cues (such as lip-reading) obtained by eye-gazing the target talker is beneficial to speech perception, especially for HI listeners [5], [6]. Therefore, when the auditory attended target switches between talkers, the listener would saccade and rotate head to direct the eye-gaze accordingly [7], [8]. Based on these studies, eye-gaze was treated as an indicator of the attended target and used to steer HAs [9]–[15]. For example, an eye-tracker-based eye-gaze selection of auditory target was reported to outperform the button-pressing-based manual selection, for both the measurement of target recalling and switching time [9]. In another study [13], an eye-gaze steering HA was found to be able to improve the speech intelligibility for HI listeners compared to non-steering one. Similar results could also be found in other studies [11], [15].

To estimate the eye-gaze, eye-tracker is the most commonly used equipment, which has high spatial resolution and low latency. However, it is usually a standalone system, requires the user's head to be stabilized and thus has low compatibility with HAs. Electrooculography (EOG) is another measurement for eye-gaze estimation [8], [14], [16], [17]. EOG signal is usually recorded by bipolar electrodes placed at the vicinity of the eyes, which measures the difference of potential evoked by eyeballs' movement, such as saccade [18]. For example, horizontal saccades would evoke prominent horizontal EOG (HEOG) and the amplitude of HEOG increases with the range of saccade [12], [19]. When the head was fixed, the hand-crafted features of HEOG (e.g., the polarity and peak value of waveform) were used to estimate the change of eye-gaze. A decision tree-based classifier with empirical thresholds was used to determine the change of eye-gaze, and the accuracy was about 65% for three pre-designed spatial-separated sources [12]. However, when the head is free of rotating, the estimation would fail since the strategy of saccade varies and the HEOG feature changes in this case. This suggested the importance of taking the head rotation into consideration for eye-gaze estimation. In a more recent study [14], a regression model was proposed to estimate the absolute eye-gaze using HEOG and head movement data recorded by an inertial measurement unit (IMU) fixed on listener's head. The regression model mapped the HEOG and IMU signals to the eye-gaze direction with a function, in which several constant parameters needed to be fitted. The estimation was reliable for the head-fixed condition but was unreliable for the head-free condition. This is likely because the fitted model was fixed and the authors assumed it applied to data of all trials. However, in face-to-face conversations, listeners usually eye-gaze at the target talker but not always with the head orienting to it [7], [15], which indicates the strategy of both saccade and head-rotation vary in realistic scenarios. These results suggested that eye-gaze estimation methods adopting the hand-crafted features or fixed-form regression function were lacked of generalization.

Besides the IMU, the neck electromyography (NEMG) recorded from the sternocleidomastoid (SCM) muscle can also be used to measure the horizontal head rotation (also called as yaw) [20], [21]. It was found that, when the head turned to right, EMG amplitudes of the left SCM increased while that of the right SCM remained unchanged, and vice versa [21]. The advantage of utilizing NEMG for head-rotation estimation is its high compatibility with HEOG equipment. However, incorporating NEMG with HEOG to estimate the eye-gaze has been rarely studied, and the feasibility still remained unknown.

In this paper, we aim to investigate the eye-gaze estimation in consideration of saccade and head-rotation behaviors, using measurement of HEOG and NEMG. Experiments with head-fixed or head-free condition are separately conducted to verify the feasibility of these measurements. Moreover, advanced classifiers, such as deep neural network (DNN) are introduced to obtain a more reliable combination of the two measurements.

## II. EXPERIMENT 1 (HEAD-FIXED)

### A. Participants and Experimental Setup

4 subjects (1 female, age range 22–25 years) with normal or corrected-normal vision participated in the experiment. All subjects were right-handed, reported no neurological illness and given informed consent approved by the Peking University Institutional Review Board.

Subjects seated in an acoustically and electrically shielded booth with their head fixed by a chinrest. In front of the chinrest at a distance of 0.4 m, a monitor was placed. In the experiment, a red dot displayed on the screen at the same horizontal level as the subject's eyes served as the stimulus. The red dot switched for every 5 s between 2 out of 5 pre-allocated positions with -45°, -30°, 0°, +30° and +45° azimuth. The subjects were instructed to eye-gaze at the red dot and switch visual attention by saccade. By pairwise combination of the 5 positions, 20 switching conditions with different starting and ending positions were presented, corresponding to 12 different eye-gaze variations ($\Delta=\pm15°, \pm30°, \pm45°, \pm60°, \pm75°$ and $\pm90°$). For each subject, there were 10 trials to be conducted, and the each trial lasted for 105 s.

### B. HEOG Recordings

Continuous HEOG data were acquired by a NeuroScan system, using a pair of bipolar electrodes located at the left and right temples, with the midforehead as GND and the nose-tip as reference. HEOG data were online bandpass filtered (0.15–100 Hz), digitized (250 Hz sampling rate), and stored for offline analysis. Along with the HEOG recording, synchronization triggers indicating the instants of red dot switching were also recorded.

### C. Signal Preprocessing and Feature Analysis

HEOG data preprocessing was performed on MATLAB. The preprocessing was similar as described in [12], [13], [17]. For each trial, 20 5-s segments of HEOG waveform corresponding to 20 switching conditions were firstly extracted according to the triggers. Each segment was clipped from 0.5 s before to 4.5 s after the switch. As suggested in [19], HEOG fluctuated slowly in time domain with its main component located in the low frequency band, the segments were then downsampled (64 Hz) and lowpass filtered (10 Hz). Finally, to ensure the amplitudes of HEOG waveforms were comparable across subjects, the segments were normalized within subject to the range [-1, 1].

Figure 1(a) shows the averaged HEOG waveforms for all conditions of eye-gaze variation. The lower standard deviation indicates that the shape of HEOG waveform is similar across subjects and trials for head-fixed condition. Across eye-gaze variations, the shapes of HEOG waveforms are also similar, except for the amplitudes. The fluctuation of waveforms all start at about 0.2 s after the switching instruction and reach maximum at about 0.4 s, and then gradually decrease to the initial level. The polarity of the peak indicates the saccade direction, i.e., positive and negative values corresponded to left and right saccades, respectively. As showed in Fig. 1(b), the relationship between eye-gaze variation and the peak value of waveform is nearly linear. The peak values increase with eye-gaze variations. These results indicate the feasibility to estimate the eye-gaze variations using HEOG for head-fixed condition.

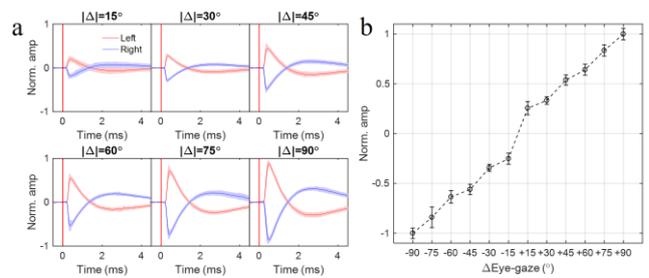

Fig. 1. (a) Normalized HEOG waveform and (b) averaged peak value of HEOG for different eye-gaze variations.

### D. Eye-gaze Estimation Algorithm

As the analyses above, peak polarity and absolute value were used as hand-crafted features of HOEG for classifying eye-gaze variations. The polarity could indicate the saccade direction (left or right) and the absolute value could indicate the amplitude of saccade. Therefore, the two features were extracted for each HEOG segment, and a support vector machine (SVM) classifier was trained and tested, taking the features as input and corresponding eye-gaze variations as output. To solve the issue that hand-crafted features are lacked of generalization, we also used DNN to automatically extract features from intact HEOG waveform. A DNN classifier with the long short-term memory (LSTM) architecture was proposed, taking HEOG waveform as input and eye-gaze variations as output. The categorical cross-entropy was used as the loss function for training. Hyperparameters of the network, such as the learning rate, number of LSTM units, and batch size were all optimized by a preliminary experiment.

The HEOG data (20 segments/trial × 10 trials/subject × 4 subjects = 800 segments) were allocated into training set (80%, 640 segments) and testing set (20%, 160 segments) with each eye-gaze variations equally distributed. To avoid the effect of dataset allocation on the classification accuracy, the dataset allocation was repeated for 100 times, as well as the training and testing sessions. The SVM model was implemented by the Scikit-learn library [22] and the LSTM model was implemented based on the Keras platform [23].

### E. Results

The overall accuracies of the SVM and LSTM classifier were 81.8±2.2% and 90.9±2.0%, respectively. The better performance of the LSTM classifier suggested the advantage of LSTM in series modeling and automatic feature extraction.

## III. EXPERIMENT 2 (HEAD-FREE)

### A. Participants and Experimental Setup

17 subjects (4 female, age range 21–28 years) with normal or corrected-normal vision participated in the experiment. All subjects were right-handed, reported no neurological illness and given informed consent approved by the Peking University Institutional Review Board.

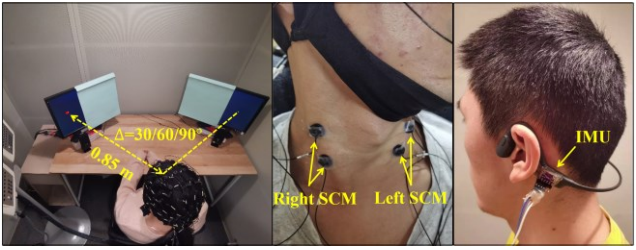

Fig. 2. The experimental environment and placement of sensors.

The experimental environment was illustrated in Fig. 2. Subjects seated in an acoustically and electrically shielded booth. In front of the subjects with a distance of 0.85 m, two monitors were symmetrically placed on the left and right with a certain of spatial separation. Similar to experiment 1, a red dot served as the stimulus. It would appear on either of the two monitors and switch its position between the two for every 5 s. Subjects were instructed to eye-gaze at the red dot and were allowed to saccade and rotate head freely during the presentation. For each trial, two fixed positions were used to display the red dot, one for each monitor, and the red dot switched between the two positions for 40 times. There were 6 potential azimuths ( -45°, -30°, -15°, +15°, +30°, and +45°) for the red dot, resulting in 6 different eye-gaze variation conditions ($\Delta=\pm30°, \pm60°$ and $\pm90°$). For each subject, 3 trials were organized according to the absolute variation of the eye-gaze variation, and each trial lasted for 205 s.

### B. HEOG, NEMG and IMU Recordings

HEOG and NEMG were also acquired by a NeuroScan system. The montage of HEOG, reference and GND electrodes were the same as in experiment 1, and the placement of other sensors was shown in Fig. 2. Referring to previous studies [21], [24], two pairs of bipolar electrodes were used to record the NEMG generated by the left and right SCM. Each pair was placed along the fibers of one SCM with a 20 mm inter-electrodes distance. HEOG and NEMG were sampled at 1 kHz. To record head-rotating related signals, a 9-axis inertial measurement unit (IMU) was used, which was attached to subject's head via a headset. The IMU worked at a sampling rate of 30 Hz. Along with HEOG, NEMG and IMU data, synchronization triggers were also recorded.

### C. Signal Preprocessing and Feature Analysis

Similar to experiment 1, for each trial, HEOG, NEMG and IMU data were spitted into 40 5-s segments. HOEG was preprocessed in the same way as in experiment 1. As suggested in [20], [21], segments of NEMG were downsampled (500 Hz), bandpass filtered (40–250 Hz) and normalized within subject. IMU data were processed using Madgwick algorithm [25] to estimate the horizontal head rotation (also called as yaw) and were also normalized.

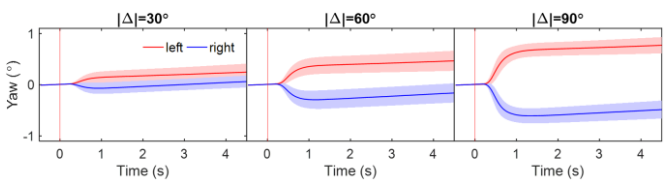

Fig. 3. Normalized head yaw for different eye-gaze variations.

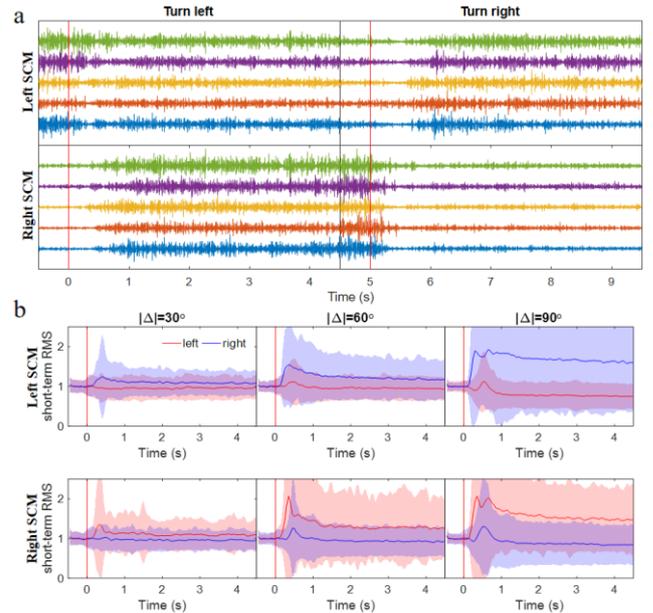

Fig. 4. (a) Individual NEMG waveforms, and (b) normalized short-term RMS value of NEMG for different eye-gaze variations.

Compared to the head-fixed experiment, HEOG had similar patterns in the head-free experiment, except that the standard deviation was larger. For IMU data, the normalized head yaw for different eye-gaze variations was showed in Fig. 3. The value of head yaw changed significantly about 0.2 s after the switching instruction and reached stability at about 1 s. The maximum value of the absolute head yaw increased with the eye-gaze variation and the polarity indicated head rotation direction. It should be mentioned that, the observed drift which might be related to hardware issue could be overlooked due to the short duration of head rotation. As shown in Fig. 4 (a), the patterns of NEMG were as expected. When the head turned left, the EMG amplitude of the left SCM decreased and that of the right SCM increased, and vice versa. To quantify such a pattern of NEMG for different eye-gaze variations, short-term RMS was calculated (0.1 s frame with 0.05 s overlap) and normalized (the first 0.5 s as the reference). As referring to Fig. 4(b), when the head turned left, the short-term RMS of the right SCM began to rise significantly about 0.1 s after the switching instruction, reached the maximum value at about 0.4 s. Similar to HEOG and head yaws, the maximum values of short-term RMS of NEMG increased with the increment of the absolute eye-gaze variation. These findings were similar to the previous study [21] and indicated the feasibility of using NEMG to measure head rotation. However, it should be noted that the standard deviations of waveform of HEOG, head yaw, and short-term RMS of NEMG were all large, which indicated that the saccade and head rotation strategy varied across subjects or trials. Therefore, it is more appropriate to estimate eye-gaze with bivariate input (e.g., HEOG+NEMG, and HEOG+IMU) which takes both saccade and head rotation into consideration, rather than univariate input (e.g., HEOG, NEMG or IMU data).

### D. Eye-gaze Estimation Algorithm

SVM was used as the feature-based classifier. The input of the SVM classifier was some of hand-crafted features of HEOG, NEMG and IMU data. Features of HEOG were the same as in experiment 1.The peaks of the short-term RMS value of NEMG of both left and right SCMs were used as features of NEMG. For IMU data, the variation of the yaw value was used as the feature. Three commonly used DNN architectures were utilized as the waveform-based classifiers, i.e., fully convolutional networks (FCN), LSTM and convolutional neural network (CNN). FCN is the most basic DNN, LSTM is good at time series modeling, and CNN has a strength in feature extraction. For these classifiers, intact waveforms of HEOG, short-term RMS of NEMG and head yaw directly served as input. Hyperparameters were also optimized by preliminary experiments. For all classifiers, both univariate input (HEOG, NEMG or IMU data) and bivariate input (HEOG+NEMG, HEOG+IMU data) were used.

The HEOG, NEMG and IMU data (40 segments/trial × 3 trials/subject × 17 subjects = 2040 segments) were allocated to training set (80%, 1632 segments) and testing set (20%, 408 segments) with each switching condition equally distributed into two sets. The dataset allocation was repeated for 50 times, as well as training and testing.

*E. Results*

Table 1 shows the classification results. For the feature-based SVM classifier, the accuracy when using univariate HEOG (67.0%) was lower than that of the head-fixed condition (81.8%). As expected, when using bivariate input, the performance was better than when using any univariate of the bivariate input (HEOG+NEMG, 68.6%, HEOG, 67.0%, NEMG, 35.1%; HEOG+IMU, 89.6%, HEOG, 67.0%, IMU, 70.5%). The results confirmed that it was insufficient to estimate eye-gaze using only saccade (HEOG) or head rotation (IMU data, NEMG) related information, and the combination of these information is significantly beneficial. It should be noted that when using NEMG instead of IMU data, the accuracy was much lower, which might due to a low quality of NEMG signals (shown by the large standard variation in Fig. 4(b)).

For the waveform-based DNN classifier, consistent with feature-based SVM classifier, the performance of bivariate input was better than when using any univariate of the bivariate input for all of the three classifiers. Comparing classification accuracies among different classifiers, SVM classifier had a similar performance to the FCN classifier, but was better than the LSTM classifier. This is likely because the LSTM is sensitive to temporal information and thus affected by the low-quality NEMG signals. However, the SVM classifier was based on hand-crafted features designed with prior-knowledge, which made it more robust to noise. The CNN classifier basically outperformed all other classifiers, probably because CNN observed the signals in a longer window at once and had more advantage at feature extracting, and thus more robust to noise.

## IV. DISCUSSION

As far as we know, this is the first study of eye-gaze estimation using HEOG and NEMG. For the head-fixed condition, the proposed model could discriminate between 12 eye-gaze variations (90% accuracy). For the head-free condition, our proposed classification model also worked reliably (waveform-based FCN: 71.8%) when using HEOG and NEMG, but the accuracy dropped dramatically when using only HEOG or NEMG data. This proved that participants' eye gazing strategies would be more complex and changeable. As suggested in [26], in natural situations, people tended to move their eyes alone for small shifts (<10°) and move their eyes and heads for larger shifts. In an auditory scene, people usually move their heads to improve speech intelligibility in noise [27] or when the target was unexpected [28]. As showed in Fig. 3 and Fig. 4 (b), the standard deviations of head yaw or NEMG RMS for each eye-gaze variation condition were large, which showed that the eye-gazing strategies were different across-subjects or even across-trials. The complexity of eye gazing strategy made it necessary to consider both saccade and head movement when estimating eye-gaze for head-free condition.

TABLE I. CLASSIFICATION RESULTS FOR HEAD-FREE CONDITION.

| Classifier / Input | Feature-based SVM | Waveform-based FCN | Waveform-based LSTM | Waveform-based CNN |
|---|---|---|---|---|
| HEOG | 67.0±1.8 | 65.3±1.6 | 49.3±4.4 | 71.0±2.1 |
| NEMG | 35.1±1.8 | 41.6±2.4 | 35.3±2.2 | 35.7±1.8 |
| IMU | 70.5±2.0 | 70.8±2.1 | 56.5±2.8 | 66.2±2.0 |
| HEOG+NEMG | 68.6±1.9 | 71.8±1.9 | 53.7±2.7 | 72.6±1.7 |
| HEOG+IMU | 89.6±1.4 | 87.9±1.5 | 76.5±3.2 | 93.3±1.3 |

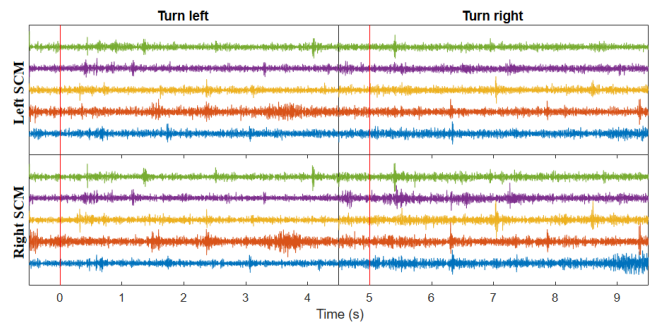

Fig. 5. Negative examples of individual NEMG waveforms.

As expected, when replacing NEMG with IMU data, the eye-gaze classification performance improved significantly for all classifiers. This could be attributed to low quality of NEMG that was not as reliable as IMU data. A negative example of individual NEMG signals was showed in Fig. 5. When head turned left or right, the value of the left or right SCM EMG was almost the same, which was unexpected. The low quality could result from the unstable attachment between electrode and the skin above the superficial SCM, since SCM would contract and stretch a lot when rotating head. This unreliability might be solved by using implanted electrodes or more recording electrodes [21].

It should be noted that the model proposed in the current study output a pre-designed eye-gaze variation by using a constant length data. The eye-gaze orientation can be calculated by summing the initial value with the variation which may bring cumulative error. In practical application, the algorithm did not know the instant when the switching happened, thus a switching detection was needed. Meanwhile, the spatial resolution was relatively low due to the limited designed eye-gaze variation classification in the experiment. More data were needed to be recorded with more eye-gaze variation conditions in the future work.

## V. CONCLUSION

In this paper, we took both saccade and head rotation into consideration for eye-gaze estimation, which was measured with HEOG and NEMG, respectively. SVM classifier was applied with hand-crafted features, and waveform-based DNN classifiers were further introduced to obtain a more reliable result. The main findings include, (1) for head-fixed condition, eye-gaze could be estimated reliably by saccade related signals (e.g., HEOG); (2) for head-free condition, an accurate eye-gaze estimation required a common using of the saccade and head rotation related signals; (3) NEMG could serve as a measurement of head-rotation, but is limited by its data quality, and (4) DNN classifiers with intact waveform as input could improve the performance of estimation than with hand-crafted feature as input. These findings could be used as a guidance for the design of eye-gaze steering HAs.

ACKNOWLEDGMENT

This work was supported by the National Natural Science Foundation of China (Grant Nos. 61771023, 12074012, and U1713217), a research funding from SONOVA, and the High-performance Computing Platform of Peking University.